\documentclass[aps,prl,twocolumn,superscriptaddress,amsmath,amssymb]{revtex4}
\usepackage{graphicx}
\usepackage{xspace}

\newcommand{\FIG}[3]{
\begin{figure}[htb]
\begin{center}
\includegraphics[width=\columnwidth]{#2}
\end{center}
\caption{#3}
\label{fig:#1}
\end{figure}}

\def\dm{\emph{Drosophila}\xspace}

\def\BCD{Bcd\xspace}
\def\HB{Hb\xspace}

\def\hb{\emph{hb}\xspace}

\def\b{B}
\def\h{H}
\def\n{n}

\def\s{\operatorname{s}}
\def\mum{\operatorname{\mu m}}
\def\nm{\operatorname{nm}}
\def\Hz{\s^{-1}}

\def\keq{K_{\rm eq}}

\def\kon{k_{\rm on}}
\def\koff{k_{\rm off}}

\def\dx{\Delta x}
\def\hc{h}
\def\db{D_{\rm \b}}

\def\fig#1{Fig.\,\ref{fig:#1}}

\def\eq#1{Eq.\,\eqref{eq:#1}}
\def\eqs#1#2{Eqs.\,\eqref{eq:#1} and \eqref{eq:#2}}

\newcommand{\e}[1]{\operatorname{e}^{#1}}

\newcommand{\Avg}[1]{\langle #1\rangle}
\newcommand{\Sinh}[1]{\sinh\negthickspace\left(\negthinspace#1\negthinspace\right)}

\newcommand{\XXX}[1]{}

\begin{document}

\date{\today}

\title{The role of spatial averaging in the precision of gene expression patterns}
\author{Thorsten Erdmann}\affiliation{FOM Institute for Atomic and Molecular Physics (AMOLF), Science Park 113, 1098 XG Amsterdam, The Netherlands}
\author{Martin Howard}\affiliation{Dept. of Computational and Systems Biology, John Innes Centre, Norwich NR4 7UH, UK}
\author{Pieter Rein ten Wolde}\affiliation{FOM Institute for Atomic and Molecular Physics (AMOLF), Science Park 113, 1098 XG Amsterdam, The Netherlands}

\begin{abstract}

During embryonic development, differentiating cells respond via gene
expression to positional cues from morphogen gradients. While gene
expression is often highly erratic, embryonic development is precise.
We show by theory and simulations that diffusion of the expressed
protein can enhance the precision of its expression domain. While
diffusion lessens the sharpness of the expression boundary, it also
reduces super-Poissonian noise by washing out bursts of gene expression.
Balancing these effects yields an optimal diffusion constant maximizing
the precision of the expression domain.

\end{abstract}

\maketitle

Embryonic development is driven by orderly, spatial patterns of gene
expression that assign each cell in the embryo its particular fate.
Experiments in recent years have vividly demonstrated that gene
expression is often highly stochastic \cite{a:ElowitzEtAl2002}. Yet,
embryonic development is exceedingly precise. A vivid example is the
Bicoid-Hunchback system in the early \dm embryo, which has become a
paradigm for understanding the formation of spatial gene-expression
patterns. Shortly after fertilization, the morphogen protein Bicoid
(\BCD) forms an exponential concentration gradient along the
anterior-posterior axis of the embryo, which provides positional
information to the differentiating nuclei. One of the target genes of
\BCD is \emph{hunchback} ({\it hb}), which is expressed in the anterior
half of the embryo.  The posterior boundary of the \hb expression domain
is very sharp: by cell cycle 13, the position of the boundary varies
only by about one nuclear spacing \cite{a:HouchmandzadehEtAl2002,
a:GregorEtAl2007b, a:HeEtAl2008b}. This precision is higher than the
best achievable precision for a time-averaging based read-out mechanism
of the \BCD gradient \cite{a:GregorEtAl2007b}. Intriguingly, the study
of Gregor {\em et al}.\ revealed that the \HB concentrations in
neighboring nuclei exhibit spatial correlations and the authors suggest
that this implies a form of spatial averaging enhancing the precision of
the posterior \HB boundary \cite{a:GregorEtAl2007b}. However, the
mechanism for spatial averaging remained unclear.

In this manuscript, we analytically and numerically study the \BCD-\HB
system. Our analysis reveals a simple, yet powerful mechanism for
spatial averaging, which is based on the diffusion of \HB itself. We
show analytically that \HB diffusion between neighboring nuclei reduces
the super-Poissonian part of the noise in its concentration, with a
factor which depends on the diffusion length of \HB and the
dimensionality of the system. In essence, diffusion reduces the noise by
washing out bursts in gene expression. This mechanism is generic, and
applies not only to any developmental system, but also to any
biochemical network in general. For example, if a signaling protein is
activated at one end of the cell and then has to diffuse to another
place to activate another system, {\em e.g}.\ the messenger ${\rm CheY}$
in bacterial chemotaxis, then our results show that the non-Poissonian
noise in the activation of the signaling protein is washed out by
diffusion; for this reason it may be beneficial to spatially separate
the in- and output of a signaling pathway. Our analysis also reveals
that while \HB diffusion reduces the noise, it also lessens the
steepness of its expression boundary \cite{a:GregorEtAl2007b}. The
interplay between these two antagonistic effects leads to an optimal
diffusion constant of $D \simeq 0.1\mum^2/\s$ that maximizes the
precision of the \hb expression domain.

We consider a minimal model of the \BCD-\HB system, which contains the
necessary ingredients for understanding the mechanism of spatial
averaging. This model includes the stochastic and cooperative activation
of \HB by \BCD in each nucleus, and the diffusion of \HB between
neighboring nuclei. To model the cooperative binding of \BCD to the \hb
promoter, we assume that \BCD proteins can bind sequentially to five
binding sites on the \hb promoter. When all five sites are occupied, the
promoter is active (activity $\n = 1$) and \HB proteins are produced
stochastically with rate $\beta$; otherwise the promoter is inactive
(activity $\n = 0$). \HB proteins are degraded stochastically with rate
$\mu$. To obtain a lower bound on the precision of the \hb expression
domain, we assume that \BCD binds to the promoter with a diffusion
limited on-rate $\kon = 4\pi\alpha\db/V = 24\alpha\db/d^3 = 8.4 \times
10^{-5}\Hz$ where $\alpha \simeq 3\nm$ is the dimension of a binding
site, $d \simeq 6.5\mum$ the diameter of a nucleus and $\db \simeq
0.32\mum^2/\s$ the \BCD diffusion constant \cite{a:GregorEtAl2007a}.
Since the on-rate is assumed to be diffusion-limited, cooperativity of
\HB activation is tuned via the \BCD off-rate $\koff(j) = a/b^j$ which
decreases with increasing number $j$ of promoter-bound \BCD. Using $a =
12.3\Hz$ and $b = 6$ the average promoter activity as function of the
number of \BCD molecules $\b$ in the nucleus approximately follows the
Hill-function
\begin{equation}\label{eq:ResponseFunction}
\Avg{\n(\b)} \simeq \b^{\hc}/(\b^{\hc} + \keq^{\hc})\,,
\end{equation}
with $\hc = 5$ and $\keq = 690$. The Hill-coefficient $\hc$ was inferred
from the relation between \HB and \BCD concentration, and  the
activation threshold $\keq$, where $\Avg{\n(\keq)} = 0.5$, is the
average number of \BCD molecules, $\Avg{B}$, in a nucleus at the \HB
domain boundary \cite{a:GregorEtAl2007b}.

To describe the formation of the spatial \HB pattern, we place $N = 64
\times 64 = 4096$ nuclei on a square lattice with spacing $\ell = 8.5
\mum$ \cite{a:GregorEtAl2007b}. \HB diffuses over the lattice with
diffusion constant $D$. With reflecting boundary conditions in one and
periodic in the other direction, our model is a cylinder which is
symmetric around the anterior-posterior $x$-axis. Differential
activation of \HB is induced by the \BCD profile
\begin{equation}\label{eq:BcdProfile}
\b(x) = \keq \e{-(x-x_t)/\lambda_{\b}},
\end{equation}
which decreases exponentially with the distance $x$ from the anterior
pole at $x = 0$. The decay length is $\lambda_{\rm \b} = 119.5 \mum$
\cite{a:GregorEtAl2007b} and the threshold position for \HB activation
is set to $x_t = 0.5L$; $L = 64\ell = 544\mum$ is the length of the
embryo.  Diffusion of \BCD between nuclei induces fluctuations in the
\BCD copy number on the time-scale $\tau_{\rm d} = \ell^2/(4\db) \simeq
10^2\s$.  Because $\tau_{\rm d}$ is much smaller than the time-scale for
promoter binding, $\kon^{-1} \simeq 10^4\s$, \BCD copy number
fluctuations are efficiently averaged out by slow binding of \BCD to the
promoter. We therefore assume that the total number of \BCD in a nucleus
is constant and given by \eq{BcdProfile}.  The state of the system is
described by the number of promoter bound \BCD and the number of \HB in
every nucleus. The state changes by reactions inside a nucleus and
diffusion of \HB between nuclei. The dynamics is described by a
reaction-diffusion master equation which we solve numerically
\cite{a:ElfEhrenberg2004}.

\FIG{Fig1}{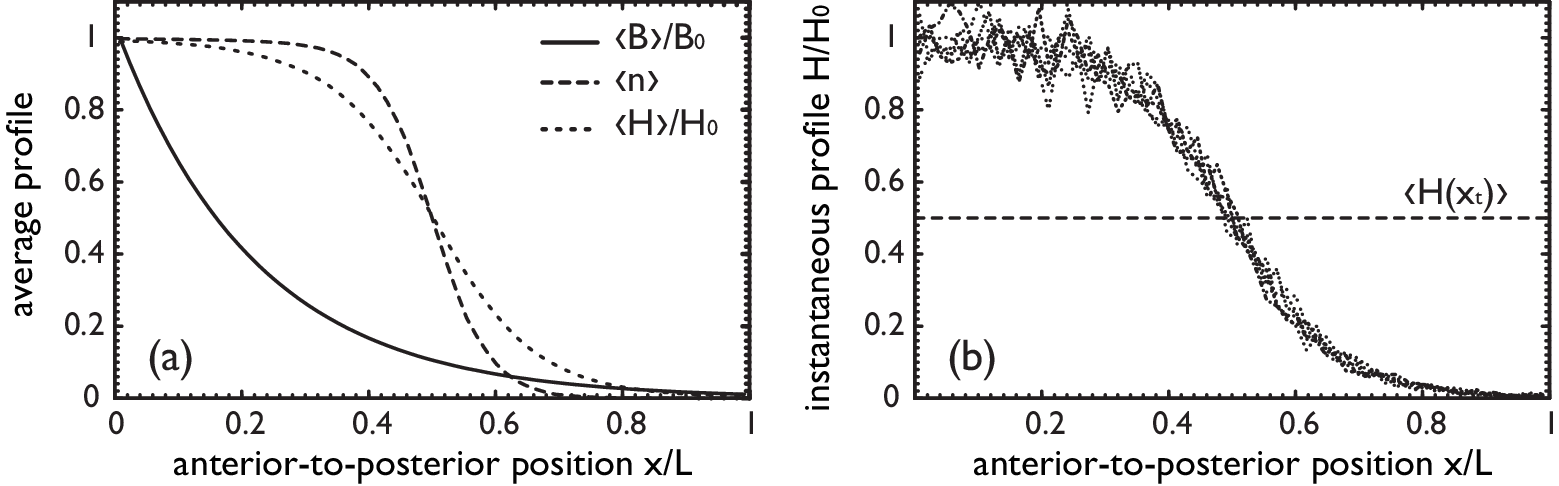}{(a) Normalized average number of free \BCD,
$\Avg{\b}/\b_0$, average promoter activity, $\Avg{\n}$, and normalized
average number of \HB, $\Avg{\h}/\h_0$, in a nucleus as function of the
position $x/L$ on the anterior-posterior axis. The \HB decay rate is
$\mu = 1.2 \times 10^{-4}\Hz$, the production rate $\beta = 500 \mu =
0.06\Hz$ and the \HB diffusion constant $D = 0.32 \mum^2/\s$.  The
off-rate of the active promoter state is $\koff(5) \simeq 1.6 \times
10^{-3}\Hz$. Normalization constants are $\b_0 = 6480$ and $\h_0 = 500$.
(b) Instantaneous normalized \HB profiles, $\h/\h_0$ at different times
in a steady-state trajectory for the same parameters as in (a). All
simulations are for the full, two-dimensional system.}

\fig{Fig1}a shows simulation results for the average number of free \BCD
molecules, $\Avg{\b}$, the average promoter activity, $\Avg{\n}$, and
the average number of \HB molecules, $\Avg{\h}$, as function of the
position $x/L$. Without \HB diffusion $\Avg{\h(x)}$ is proportional to
$\Avg{\n(x)}$. For a finite \HB diffusion constant, however, the shape
of the \HB profile is determined not only by $\Avg{\n(x)}$, but also by
the diffusion length of \HB, $\lambda = \sqrt{D/\mu}$---with increasing
$\lambda$ the profile becomes less steep. \fig{Fig1}b shows
instantaneous \HB profiles at different times. Fluctuations of promoter
activity induce large fluctuations of $\h$. These in turn lead to an
uncertainty $\dx$ in the position at which $\h$ crosses the threshold
$\Avg{\h(x_t)} = 0.5 \beta/\mu$.

\FIG{Fig2}{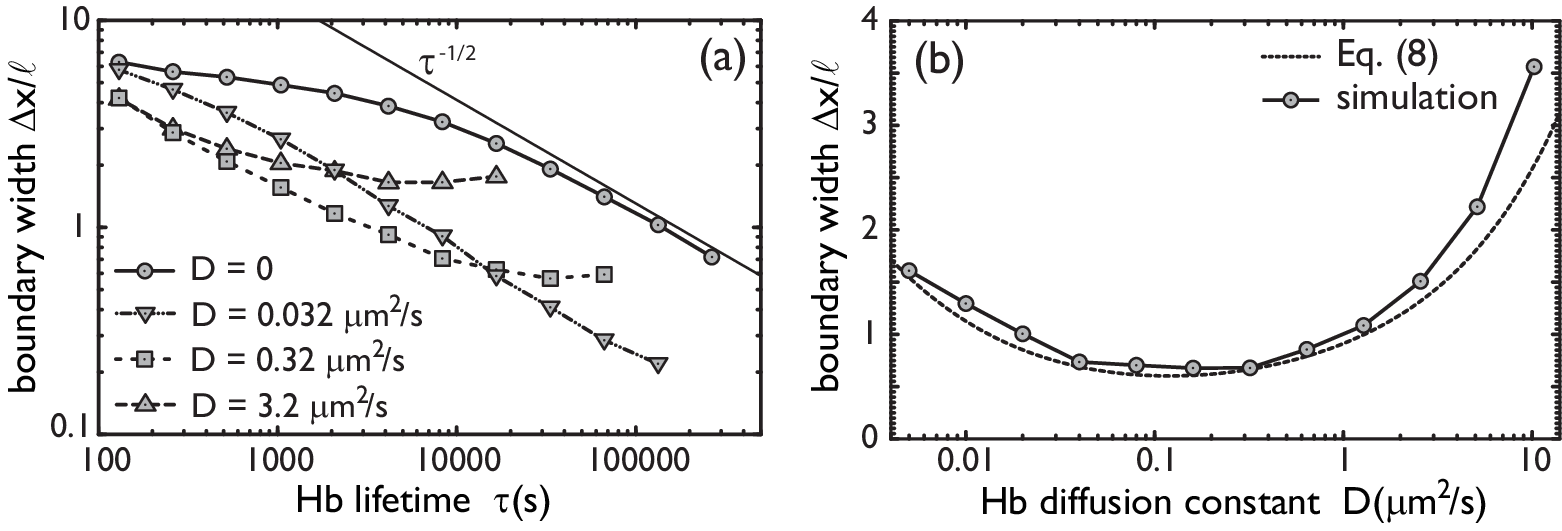}{(a) Simulation results for the \HB boundary width
$\dx$ as a function of the \HB lifetime $\tau = \mu^{-1}$ for a constant
production rate $\beta = 0.06\Hz$ and different diffusion constants $D$.
The solid line scales as $\tau^{-1/2}$. (b) Simulation results for $\dx$
as a function of $D$ and for $\beta$ and $\mu$ as in \fig{Fig1}.  The
simulation results are compared to \eq{BoundaryWidth} using
\eqs{HbGradient}{VarianceReservoir}. All simulations are for the full,
two-dimensional system.}

We now consider the \HB boundary width when the \HB diffusion constant
$D$ is zero. \fig{Fig2}a shows that this decreases with increasing
lifetime of the \HB protein. To understand this behavior, we note that
the boundary width is, to a good approximation, given by
\begin{equation}\label{eq:BoundaryWidth}
\dx = \frac{\sigma(x_t)}{|\Avg{\h(x_t)}'|},
\end{equation}
where $\sigma(x_t)$ is the standard deviation of the \HB copy number and
$|\Avg{\h(x_t)}'|$ is the magnitude of the \HB gradient at the boundary
position $x_t$ \cite{a:TostevinEtAl2007, a:GregorEtAl2007b}.  When the
\HB lifetime $\tau$ is much longer than the time-scale $\koff(5)^{-1}$
of the promoter state fluctuations, then the noise in \HB copy number
becomes Poissonian and the variance is given by the mean: $\sigma^2(x_t)
= \Avg{\h(x_t)} = \Avg{n(x_t)} \beta/\mu = 0.5 \beta\tau$. On the other
hand, the steepness of the boundary, ${|\Avg{\h(x_t)}'|}$, increases
linearly with $\tau$ when $D=0$ and the synthesis rate $\beta$ is kept
constant. \eq{BoundaryWidth} thus predicts that the boundary width
$\Delta x$ decreases as $\tau^{-1/2}$ for large $\tau \gg
\koff(5)^{-1}$, which is indeed observed in the simulations (see
\fig{Fig2}a). Hence, the \hb expression boundary could be made
arbitrarily precise if the lifetime of the \HB protein could be
increased indefinitely.

In practice the averaging time cannot be made arbitrarily long
\cite{a:GregorEtAl2007b}.  Ultimately, it is limited by the nuclear
division time $\tau_{\rm nd}$, which at cell cycle 13 is roughly 30
minutes. \fig{Fig2}a shows that for $\tau \simeq \tau_{\rm nd}$ and
vanishing \HB diffusion constant $D = 0$, the boundary width would be
$\dx \simeq 4$ nuclei, which is larger than the precision measured
experimentally \cite{a:GregorEtAl2007b}. The limited averaging time thus
puts strong constraints on the precision that can be achieved via the
mechanism of time-averaging alone.

\fig{Fig2}a reveals, however, that the precision of the \hb expression
domain can be enhanced significantly by increasing the \HB diffusion
constant to a finite value. This may seem surprising, since the
steepness of the boundary, $|\Avg{\h(x_t)}'|$, decreases with increasing
\HB diffusion constant (\fig{Fig1}) and this---as \eq{BoundaryWidth}
shows---tends to increase the boundary width. However, increasing the
\HB diffusion constant also reduces the noise in the \HB copy number.
This is the mechanism of spatial averaging, which we now study
analytically.

\FIG{Fig3}{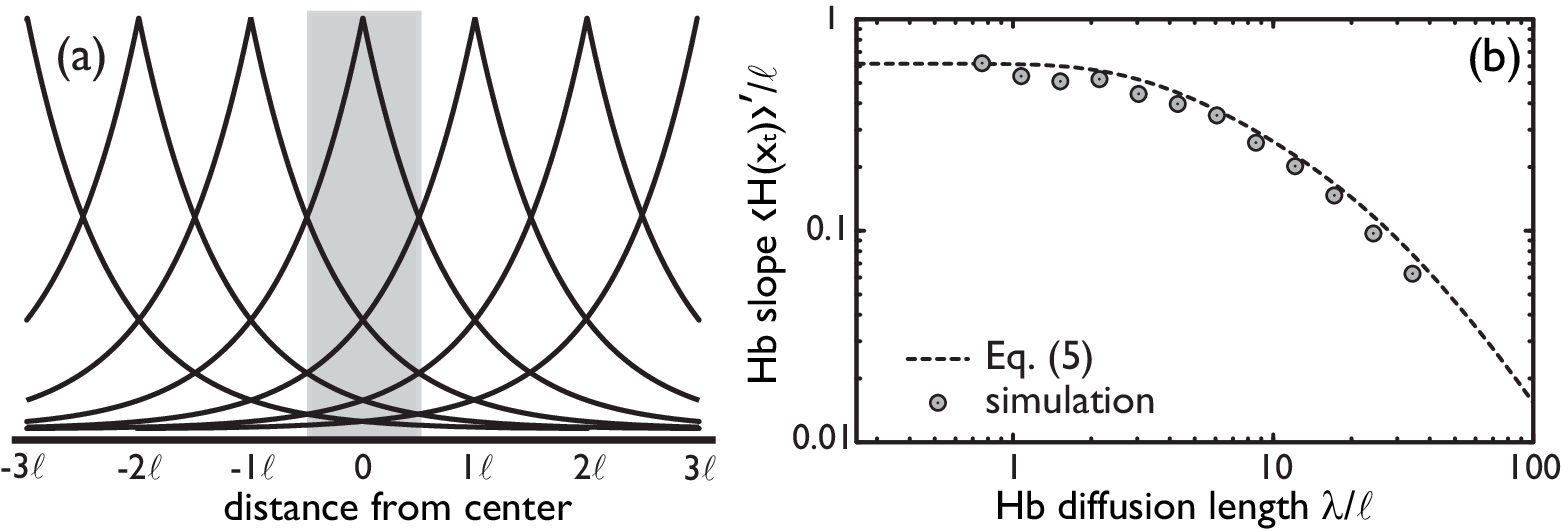}{(a) \HB is produced in nuclei and diffuses
away from these \HB point sources. The stationary number of \HB in a
given nucleus (shaded area) is the sum over contributions from all
nuclei. The size of each contribution reduces exponentially with the
distance from its source. (b) Slope $\Avg{\h(x_t)}'$ of the \HB profile
at the boundary $x_t$ as function of the \HB diffusion length $\lambda =
\sqrt{D/\mu}$. All lengths are in units of the lattice constant $\ell =
8.5\mum$. Analytical results from \eq{HbGradient} (dashed line) are
compared to simulation results (symbols) for the full, two-dimensional
system.}

To elucidate the mechanism of spatial averaging, we first analyze how
the steepness of the \HB boundary, $|\Avg{\h(x_t)}'|$, depends on the
\HB diffusion length $\lambda$, and then how the noise in the \HB copy
number at the boundary, $\sigma(x_t)$, depends on $\lambda$ (see
\eq{BoundaryWidth}). To calculate the \HB profile, we note that a
nucleus at a position $x_i$ produces \HB with an average rate $\beta
\Avg{\n(x_i)}$ (\fig{Fig3}a). In one dimension, the steady state number
of \HB molecules in a nucleus of size $\ell$ at $x$ is approximately
\begin{equation}\label{eq:PointSourceProfile}
\Avg{\h(x)}_{\rm ps} \approx \frac{\beta\Avg{\n(x_i)}/\mu}{2(\lambda/\ell)}\e{-|x-x_i|/\lambda}\,.
\end{equation}
The total \HB profile is the sum of $\Avg{\h(x)}_{\rm ps}$ over all
point sources $x_i = 0, \ell, 2\ell, \dots$. To do the sum, we
approximate the activity $\Avg{\n(x_i)}$ from
\eqs{ResponseFunction}{BcdProfile} by a piecewise linear function which
behaves as $\Avg{\n(x_i)} = 0.5 + m(x_t - x_i)$ around the boundary for
$x_t - 1/2m < x_i < x_t + 1/2m$ and which is unity below and zero above
the boundary. The slope $-m = -\hc / 4\lambda_{\rm \b}$ is the
derivative of $\Avg{\n(x_i)}$ at $x_i = x_t$. At the boundary $x_t$ the
slope of $\Avg{\h(x)}$ is
\begin{equation}\label{eq:HbGradient}
\Avg{\h(x_t)}' = m\frac{\beta}{\mu}\left\{1 - \e{-\frac{1}{2m\lambda}} - \Sinh{\frac{1}{2m\lambda}}\e{-\frac{2x_t}{\lambda}}\right\}.
\end{equation}
It increases linearly with the number of \HB molecules $\beta/\mu$ and
decreases as the \HB diffusion length increases. Due to the rotational
symmetry, the slope in the full, two-dimensional system is also given by
\eq{HbGradient}. \fig{Fig3}b plots \eq{HbGradient} as function of
$\lambda$ and compares the approximation with two-dimensional simulation
results. It is seen that the agreement is very good.

\FIG{Fig4}{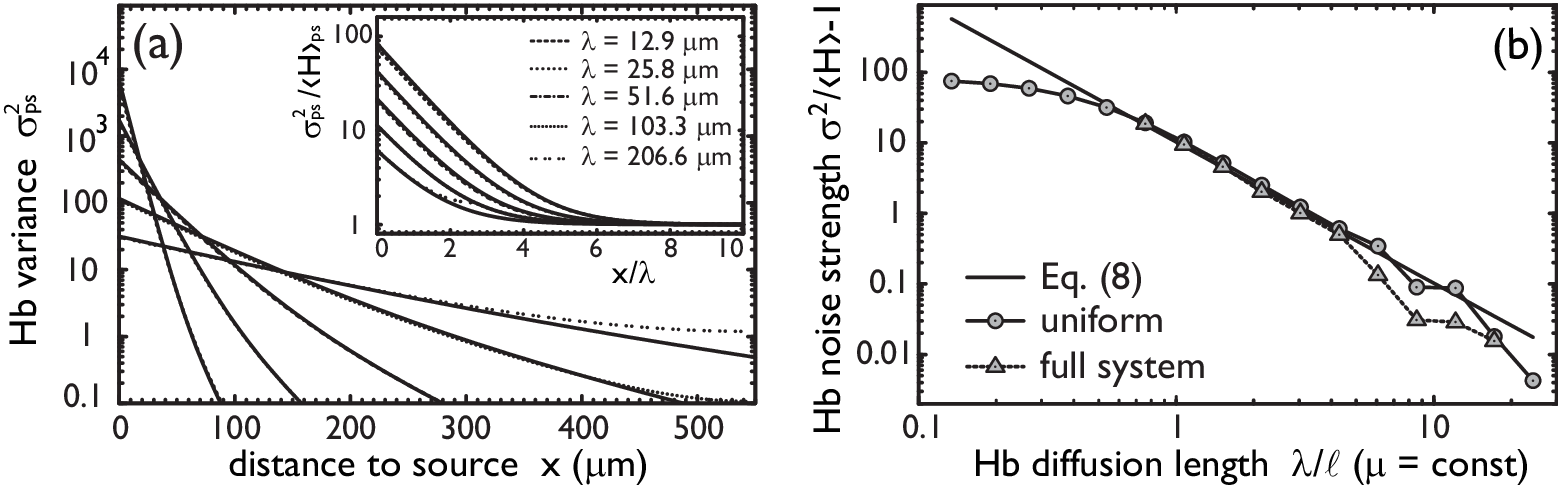}{(a) Variance $\sigma_{\rm ps}^2(x)$ of \HB as
function of the distance $x$ from a point source at $x = 0$ in
one-dimensional space for increasing \HB diffusion length $\lambda$.
\eq{VarianceLimit} is compared to simulation results for $\mu = 1.2
\times 10^{-2}\Hz$ and $\beta = 500\mu = 6\Hz$, {\em i.e}.\ for fast \HB
dynamics. While a promoter is active, $\beta/\koff(5) \simeq 3800$ \HB
molecules are produced on average. The \HB diffusion length is varied
through the diffusion coefficient $D$. The inset shows the \HB noise
strength $\sigma_{\rm ps}^2(x/\lambda) / \Avg{\h(x/\lambda)}_{\rm ps}$
as function of distance normalized by the diffusion length. (b)
Non-Poissonian part of the \HB noise strength, $\sigma^2/\Avg{\h} - 1$,
for a uniform system with $N = 32 \times 32$ nuclei on a square lattice
as function of $\lambda/\ell$. Decay and production rate of \HB are
constant, while $D$ is varied; for other parameter values, see
\fig{Fig1}. The average promoter activity is $\Avg{\n} = 0.5$.
Simulation results for the uniform system (circles) are compared to
\eq{VarianceReservoir} with the two-dimensional value $N(\lambda) = 8
(\lambda/\ell)^2$ (solid line).  The value $\sigma_0^2/\Avg{\h} - 1 =
82.4$ (see \eq{VarianceReservoir}) in an isolated volume has been
calculated numerically. These results are compared to the \HB noise
strength for the full, two-dimensional system (see \fig{Fig1}) in a
nucleus at the boundary with $\Avg{\n} = 0.5$ (triangles).}

Next, we study the effect of \HB diffusion on the variance in the \HB
copy number, $\sigma^2$ (see \eq{BoundaryWidth}). For clarity, we first
do the calculations for one-dimensional diffusion and extend our
approximation to general dimensions afterwards. To compute $\sigma^2$,
we exploit the observation made above that each nucleus acts as a point
source of \HB (\fig{Fig3}a), and that the expression of \HB in each
nucleus is an independent stochastic process. We denote the variance of
the \HB copy number in a nucleus of size $\ell$ at a distance $x$ from
the point source as $\sigma_{\rm ps}^2(x)$. In a uniform space, the
total \HB variance in a nucleus is the sum $\sigma^2 = \sum_i
\sigma_{\rm ps}^2(x_i)$ over point sources at all $x_i$. To calculate
$\sigma^2_{\rm ps}(x)$, we assume that the \HB dynamics is fast on the
time-scale of promoter switching, such that the \HB concentration
switches between zero and $\Avg{\h(x)}_{\rm ps} / \Avg{\n}$, with
$\Avg{\h(x)}_{\rm ps}$ given by \eq{PointSourceProfile}. In this limit,
\begin{equation}\label{eq:VarianceLimit}
\sigma_{\rm ps}^2(x) \approx \Avg{\h(x)}_{\rm ps} + \Avg{\h(x)}_{\rm ps}^2\frac{\Avg{1-\n}}{\Avg{\n}},
\end{equation}
where the second term is due to promoter switching with variance $\Avg{1
- n}\Avg{n}$, and the first term describes the Poisson noise coming from
the production, diffusion and decay of \HB when the promoter is active.

\fig{Fig4}a shows $\sigma^2_{\rm ps}(x)$ for different values of the
diffusion length $\lambda$. For small $\lambda$, $\sigma_{\rm ps}^2(x)$
is large for small $x$, but reduces quickly with increasing $x$. Hence,
the variance at a given nucleus is determined by few nuclei in the
immediate neighborhood but each individual contribution is large. For
increasing $\lambda$ the number of nuclei contributing to the variance
increases, but the individual contributions of nearby nuclei are
smaller; this is because diffusion washes out the bursts of \HB
production at the source. To see which of the two opposing effects
dominates, we sum \eq{VarianceLimit} over $x_i$ to obtain the total
variance:
\begin{equation}\label{eq:VarianceAverage}
\sigma^2 = \Avg{\h} + \frac{\Avg{\h}^2}{4(\lambda/\ell)}\frac{\Avg{1-\n}}{\Avg{\n}}\,.
\end{equation}
Compared to the variance in an isolated nucleus, the overall effect of
diffusion in a one-dimensional, uniform space is a reduction of the
non-Poissonian noise by $(4\lambda/\ell)^{-1}$. The Poissonian part
cannot be reduced because diffusion itself is Poissonian. Indeed,
spatial averaging can reduce the effect of noise in gene expression, but
only if this is super-Poissonian. Simulations also allow us to compute
$\sigma_{\rm ps}^2(x)$ for slower \HB dynamics. In this case, the
non-Poissonian part of the variance is smaller. However, the qualitative
dependence of $x$ remains the same: the non-Poissonian part is washed
out by diffusion over a distance proportional to $\lambda$ (inset
\fig{Fig4}a).

The finite range of non-Poissonian fluctuations implies the following
simple model. $N$ nuclei contribute equally to a well-stirred reservoir
containing $M$ \HB molecules, with average $\Avg{M} = N\Avg{\h}$ and
variance $\sigma^2_M = N \sigma_0^2$, where $\sigma^2_0$ is the noise of
$\h$ in a nucleus if there was no \HB diffusion. Each nucleus then
samples \HB molecules in a binomial fashion from this reservoir. The
noise in each nucleus is formally given by $\sigma^2 = E[V[H|M]] +
V[E[H|M]]$, where $E[X]$ denotes the expectation and $V[X]$ the variance
of $X$. The first term describes the noise in the sampling process, and
is given by $E[V[H|M]]=\Avg{\h}(1-N^{-1})$. The second term describes
the variance in $\h$ due to fluctuations in the reservoir, and is given
by $V[E[H|M]] = \sigma_0^2/N$. Hence,
\begin{equation}\label{eq:VarianceReservoir}
\sigma^2 = \Avg{\h} + \frac{\sigma_0^2 - \Avg{\h}}{N}\,.
\end{equation}
This simple expression elucidates that it is the super-Poissonian part
of the noise, $\sigma_0^2-\Avg{H}$, which is reduced by spatial
averaging. Since $N$ scales with the dimensionality $d$ as $N \sim
\lambda^d$, this reduction becomes more efficient in higher dimensions.
Interestingly, this expression also reveals that if \hb expression would
be {\em sub}-Poissonian \cite{a:PedrazaPaulsson2008}, diffusion would
\emph{increase} the noise.

\fig{Fig4}b shows the non-Poissonian part of the variance in a uniform,
two-dimensional system as function of the diffusion length; the exact,
numerical result is compared to the prediction of \eq{VarianceReservoir}
with $N$ chosen to be the number of nuclei within a distance $2\lambda$
of a given nucleus along the edges of a square lattice, $N(\lambda) =
8(\lambda/\ell)^2$. For sufficiently large $\lambda$ for which
$N(\lambda) > 1$, the approximation is excellent. \fig{Fig4}b also
compares the results for the uniform system to those of the full, {\em
non-uniform} system in which a \BCD gradient activates \HB. Clearly, the
well-stirred approximation describes the variance at the \HB boundary
very well; only for $\lambda > 5\ell$ is the variance at the boundary
significantly influenced by nuclei far away from the boundary with
smaller variance.

Finally, we can combine \eqs{HbGradient}{VarianceReservoir} with
\eq{BoundaryWidth} to predict the boundary width $\dx$. \fig{Fig2}b
shows that the prediction agrees very well with the simulations. This
figure also shows that the two antagonistic effects of \HB
diffusion---reducing the slope but also the variance of the \HB
concentration---lead to a diffusion constant that optimizes the boundary
width for a fixed \HB lifetime. Interestingly, this minimal width is
less than one nuclear spacing, as found experimentally
\cite{a:GregorEtAl2007b}. This suggests that for achieving the necessary
precision, mechanisms based on multiple gradients or interactions
between multiple \BCD targets  are not required, although these may
provide robustness against embryo-to-embryo variations
\cite{a:HowardTenWolde2005, a:ManuEtAl2009a, a:ManuEtAl2009b}. Since the
exact mechanism for \HB transport is unknown, it is tempting to
speculate that the precision achieved via spatial averaging could be
increased further by separating the two effects of \HB diffusion using
an anisotropic transport of \HB: slow transport along the
anterior-posterior axis would allow for steep spatial profiles, while
fast diffusion along the perpendicular direction would allow for
effective spatial averaging. The benefit of anisotropic transport could
imply the existence of active mechanisms for \HB transport.

Spatial averaging can only be beneficial when the noise in \HB
production has a super-Poissonian component. The observation of spatial
correlations in \HB by Gregor \emph{et al}.\ indicates that this is the
case \cite{a:GregorEtAl2007b}. The question arises whether bursts in
gene expression are inevitable and spatial averaging a prerequisite for
achieving a precise \hb expression domain. The \HB lifetime is limited
by the nuclear division time, while the time-scale for promoter state
fluctuations is limited by the diffusion of \BCD. The magnitude of the
bursts could be reduced by reducing the promoter strength, but this
would lower the \HB copy number and hence the steepness of the \HB
boundary; the net result would, in fact, be a {\em decrease} of its
precision. In a model with explicit \hb mRNA, the effect of promoter
state fluctuations could be alleviated by reducing the transcription
rate.  However, to achieve a sufficiently steep and precise \HB boundary
with a low number of \hb mRNA, the translation rate had to be increased
such that \HB production becomes super-Poissonian again. It thus appears
that spatial averaging is a fundamental mechanism for generating precise
patterns of gene expression. Clearly, it will be of interest to study
how other modes of transport, {\em e.g.}\ sub-diffusion or active
transport, affect the mechanism of spatial averaging.

We thank Filipe Tostevin for a critical reading of the manuscript. This
work is supported by FOM/NWO (TE, PRtW) and The Royal Society (MH).

%\bibliographystyle{apsrev}
%\bibliography{prl}

\begin{thebibliography}{11}
\expandafter\ifx\csname natexlab\endcsname\relax\def\natexlab#1{#1}\fi
\expandafter\ifx\csname bibnamefont\endcsname\relax
  \def\bibnamefont#1{#1}\fi
\expandafter\ifx\csname bibfnamefont\endcsname\relax
  \def\bibfnamefont#1{#1}\fi
\expandafter\ifx\csname citenamefont\endcsname\relax
  \def\citenamefont#1{#1}\fi
\expandafter\ifx\csname url\endcsname\relax
  \def\url#1{\texttt{#1}}\fi
\expandafter\ifx\csname urlprefix\endcsname\relax\def\urlprefix{URL }\fi
\providecommand{\bibinfo}[2]{#2}
\providecommand{\eprint}[2][]{\url{#2}}

\bibitem[{\citenamefont{Elowitz et~al.}(2002)\citenamefont{Elowitz, Levine,
  Siggia, and Swain}}]{a:ElowitzEtAl2002}
\bibinfo{author}{\bibfnamefont{M.~B.} \bibnamefont{Elowitz}},
  \bibinfo{author}{\bibfnamefont{A.~J.} \bibnamefont{Levine}},
  \bibinfo{author}{\bibfnamefont{E.~D.} \bibnamefont{Siggia}},
  \bibnamefont{and} \bibinfo{author}{\bibfnamefont{P.~S.} \bibnamefont{Swain}},
  \bibinfo{journal}{Science} \textbf{\bibinfo{volume}{297}},
  \bibinfo{pages}{1183} (\bibinfo{year}{2002}).

\bibitem[{\citenamefont{Houchmandzadeh
  et~al.}(2002)\citenamefont{Houchmandzadeh, Wieschaus, and
  Leibler}}]{a:HouchmandzadehEtAl2002}
\bibinfo{author}{\bibfnamefont{B.}~\bibnamefont{Houchmandzadeh}},
  \bibinfo{author}{\bibfnamefont{E.}~\bibnamefont{Wieschaus}},
  \bibnamefont{and} \bibinfo{author}{\bibfnamefont{S.}~\bibnamefont{Leibler}},
  \bibinfo{journal}{Nature} \textbf{\bibinfo{volume}{415}},
  \bibinfo{pages}{798} (\bibinfo{year}{2002}).

\bibitem[{\citenamefont{Gregor et~al.}(2007{\natexlab{a}})\citenamefont{Gregor,
  Tank, Wieschaus, and Bialek}}]{a:GregorEtAl2007b}
\bibinfo{author}{\bibfnamefont{T.}~\bibnamefont{Gregor}},
  \bibinfo{author}{\bibfnamefont{D.~W.} \bibnamefont{Tank}},
  \bibinfo{author}{\bibfnamefont{E.~F.} \bibnamefont{Wieschaus}},
  \bibnamefont{and} \bibinfo{author}{\bibfnamefont{W.}~\bibnamefont{Bialek}},
  \bibinfo{journal}{Cell} \textbf{\bibinfo{volume}{130}}, \bibinfo{pages}{153}
  (\bibinfo{year}{2007}{\natexlab{a}}).

\bibitem[{\citenamefont{He et~al.}(2008)\citenamefont{He, Wen, Deng, Lin, Lu,
  Jiao, and Ma}}]{a:HeEtAl2008b}
\bibinfo{author}{\bibfnamefont{F.}~\bibnamefont{He}},
  \bibinfo{author}{\bibfnamefont{Y.}~\bibnamefont{Wen}},
  \bibinfo{author}{\bibfnamefont{J.}~\bibnamefont{Deng}},
  \bibinfo{author}{\bibfnamefont{X.}~\bibnamefont{Lin}},
  \bibinfo{author}{\bibfnamefont{L.~J.} \bibnamefont{Lu}},
  \bibinfo{author}{\bibfnamefont{R.}~\bibnamefont{Jiao}}, \bibnamefont{and}
  \bibinfo{author}{\bibfnamefont{J.}~\bibnamefont{Ma}},
  \bibinfo{journal}{Developmental Cell} \textbf{\bibinfo{volume}{15}},
  \bibinfo{pages}{558} (\bibinfo{year}{2008}).

\bibitem[{\citenamefont{Gregor et~al.}(2007{\natexlab{b}})\citenamefont{Gregor,
  Wieschaus, McGregor, Bialek, and Tank}}]{a:GregorEtAl2007a}
\bibinfo{author}{\bibfnamefont{T.}~\bibnamefont{Gregor}},
  \bibinfo{author}{\bibfnamefont{E.~F.} \bibnamefont{Wieschaus}},
  \bibinfo{author}{\bibfnamefont{A.~P.} \bibnamefont{McGregor}},
  \bibinfo{author}{\bibfnamefont{W.}~\bibnamefont{Bialek}}, \bibnamefont{and}
  \bibinfo{author}{\bibfnamefont{D.~W.} \bibnamefont{Tank}},
  \bibinfo{journal}{Cell} \textbf{\bibinfo{volume}{130}}, \bibinfo{pages}{141}
  (\bibinfo{year}{2007}{\natexlab{b}}).

\bibitem[{\citenamefont{Elf and Ehrenberg}(2004)}]{a:ElfEhrenberg2004}
\bibinfo{author}{\bibfnamefont{J.}~\bibnamefont{Elf}} \bibnamefont{and}
  \bibinfo{author}{\bibfnamefont{M.}~\bibnamefont{Ehrenberg}},
  \bibinfo{journal}{Systems Biology} \textbf{\bibinfo{volume}{1}},
  \bibinfo{pages}{230} (\bibinfo{year}{2004}).

\bibitem[{\citenamefont{Tostevin et~al.}(2007)\citenamefont{Tostevin, ten
  Wolde, and Howard}}]{a:TostevinEtAl2007}
\bibinfo{author}{\bibfnamefont{F.}~\bibnamefont{Tostevin}},
  \bibinfo{author}{\bibfnamefont{P.~R.} \bibnamefont{ten Wolde}},
  \bibnamefont{and} \bibinfo{author}{\bibfnamefont{M.}~\bibnamefont{Howard}},
  \bibinfo{journal}{PLoS Computational Biology} \textbf{\bibinfo{volume}{3}},
  \bibinfo{pages}{763} (\bibinfo{year}{2007}).

\bibitem[{\citenamefont{Pedraza and Paulsson}(2008)}]{a:PedrazaPaulsson2008}
\bibinfo{author}{\bibfnamefont{J.~M.} \bibnamefont{Pedraza}} \bibnamefont{and}
  \bibinfo{author}{\bibfnamefont{J.}~\bibnamefont{Paulsson}},
  \bibinfo{journal}{Science} \textbf{\bibinfo{volume}{319}},
  \bibinfo{pages}{339} (\bibinfo{year}{2008}).

\bibitem[{\citenamefont{Howard and ten Wolde}(2005)}]{a:HowardTenWolde2005}
\bibinfo{author}{\bibfnamefont{M.}~\bibnamefont{Howard}} \bibnamefont{and}
  \bibinfo{author}{\bibfnamefont{P.~R.} \bibnamefont{ten Wolde}},
  \bibinfo{journal}{Physical Review Letters} \textbf{\bibinfo{volume}{95}},
  \bibinfo{pages}{208103} (\bibinfo{year}{2005}).

\bibitem[{\citenamefont{Manu et~al.}(2009{\natexlab{a}})\citenamefont{Manu,
  Surkova, Spirov, Gursky, Janssens, Kim, Radulescu, Vanario-Alonso, Sharp,
  Samsonova et~al.}}]{a:ManuEtAl2009a}
\bibinfo{author}{\bibnamefont{Manu}},
  \bibinfo{author}{\bibfnamefont{S.}~\bibnamefont{Surkova}},
  \bibinfo{author}{\bibfnamefont{A.~V.} \bibnamefont{Spirov}},
  \bibinfo{author}{\bibfnamefont{V.~V.} \bibnamefont{Gursky}},
  \bibinfo{author}{\bibfnamefont{H.}~\bibnamefont{Janssens}},
  \bibinfo{author}{\bibfnamefont{A.-R.} \bibnamefont{Kim}},
  \bibinfo{author}{\bibfnamefont{O.}~\bibnamefont{Radulescu}},
  \bibinfo{author}{\bibfnamefont{C.~E.} \bibnamefont{Vanario-Alonso}},
  \bibinfo{author}{\bibfnamefont{D.~H.} \bibnamefont{Sharp}},
  \bibinfo{author}{\bibfnamefont{M.}~\bibnamefont{Samsonova}},
  \bibnamefont{et~al.}, \bibinfo{journal}{PLoS Biology}
  \textbf{\bibinfo{volume}{7}}, \bibinfo{pages}{e1000049}
  (\bibinfo{year}{2009}{\natexlab{a}}).

\bibitem[{\citenamefont{Manu et~al.}(2009{\natexlab{b}})\citenamefont{Manu,
  Surkova, Spirov, Gursky, Janssens, Kim, Radulescu, Vanario-Alonso, Sharp,
  Samsonova et~al.}}]{a:ManuEtAl2009b}
\bibinfo{author}{\bibnamefont{Manu}},
  \bibinfo{author}{\bibfnamefont{S.}~\bibnamefont{Surkova}},
  \bibinfo{author}{\bibfnamefont{A.~V.} \bibnamefont{Spirov}},
  \bibinfo{author}{\bibfnamefont{V.~V.} \bibnamefont{Gursky}},
  \bibinfo{author}{\bibfnamefont{H.}~\bibnamefont{Janssens}},
  \bibinfo{author}{\bibfnamefont{A.-R.} \bibnamefont{Kim}},
  \bibinfo{author}{\bibfnamefont{O.}~\bibnamefont{Radulescu}},
  \bibinfo{author}{\bibfnamefont{C.~E.} \bibnamefont{Vanario-Alonso}},
  \bibinfo{author}{\bibfnamefont{D.~H.} \bibnamefont{Sharp}},
  \bibinfo{author}{\bibfnamefont{M.}~\bibnamefont{Samsonova}},
  \bibnamefont{et~al.}, \bibinfo{journal}{PLoS Computational Biology}
  \textbf{\bibinfo{volume}{5}}, \bibinfo{pages}{e1000303}
  (\bibinfo{year}{2009}{\natexlab{b}}).

\end{thebibliography}

\end{document}